# A ROUTING MECHANISM BASED ON SOCIAL NETWORKS AND BETWEENNESS CENTRALITY IN DELAY-TOLERANT NETWORKS

Zhang Huijuan[1] and Liu Kai[2]

School of Software Engineering, Tongji University, Shanghai, China

## ABSTRACT

*With the growing popularity of mobile smart devices, the existing networks are unable to meet the requirement of many complex scenarios; current network architectures and protocols do not work well with the network with high latency and frequent disconnections. To improve the performance of these networks some scholars opened up a new research field, delay-tolerant networks, in which one of the important research subjects is the forwarding and routing mechanism of data packets. This paper presents a routing scheme based on social networks owing to the fact that nodes in computer networks and social networks have high behavioural similarity. To further improve efficiency this paper also suggests a mechanism, which is the improved version of an existing betweenness centrality based routing algorithm [1]. The experiments showed that the proposed scheme has better performance than the existing friendship routing algorithms.*

## 1. INTRODUCTION

The existing TCP/IP protocols guarantee stable and efficient data transmission, based on the following premises [2]:

1) There is at least one end-to-end path between data source and destination.

2) The maximum round-trip time (RTT) between any node in the network is not excessive.

3) The packet loss rate is relatively low.

Unfortunately, a class of challenged network which don't meet the above conditions may not able to be served well by the current TCP/IP model. For instance, some networks may not have sufficient energy supply, some nodes in the network may not have enough storage space to relay the data packets, data source may not have stable end-to-end paths to its peers, or the networks with latency which always cause unacceptable delays. All these examples bring new challenges to current network technology.

To solve such kind of problems, K. Fall demonstrates a new network architecture, delay-tolerant network (DTN) at SIGCOMM 2003 [3]. Delay-tolerant networks refer to the networks which cannot have a stable end-to-end path for various reasons; the status of the networks may even appear as disconnected most of the time. Compared with the traditional network architecture, DTN has such characteristics as self-organized and dynamic topology. Therefore, an efficient store-carry-forward routing scheme is essential to DTNs.

Many studies have shown that the movements of nodes in social networks and DTNs are based on similar motion patterns; nodes tent to have mobility patterns influenced by their social relationships or social behaviour [4]. It may be possible to improve DTN routing efficiency by increasing the priorities of some nodes that are much socially related.

In most sociality-based DTNs, nodes meet other nodes then store the contact information. Some existing methods analyse social relationship of nodes based on such kind of contact history. E. Bulut et al. introduced a new metric to detect the quality of friendships of each node accurately [5]. They calculate the link weights between each pair of nodes then construct a friendship network based on the link weights. A sociality-based routing scheme, Friendship Routing, which utilizes such friendship network to make the forwarding decisions of data packets. C. M. Kim et al. present a routing strategy which uses expanded ego-network betweenness centrality when a relay node is selected [1].

In this paper, we propose a routing scheme in which each node determines the data packet's next hop based on the contact history. Endpoint-biased expanded ego-network betweenness centrality is used to enhance the overall routing efficiency. We also demonstrated the proposed scheme has better performance than the existing friendship routing algorithm.

The rest of this paper is organized as follows. Section 2 explains how to construct the social network and the detailed definition of betweenness centralities (Primitive, Expanded Ego; Section 3 presents the proposed routing scheme; Section 4 shows the simulation analysis and the Section 5 concludes the paper.

## 2. VIRTUAL SOCIAL NETWORK CONSTRUCTION

We suppose the nodes in DTNs keep moving in a long time frame. Here we assume the transmission range of a node is $r$. A node i *encounters* another node j when these two nodes are within transmission range $r$ from each other; node i receives a first hello message broadcasted by node j at the moment. If node i and node j stay within the transmission range of each other, both will get a periodic hello message from its peer. We consider node i leaves node j if node i does not hear a predefined number of hello messages from node j continuously. If node i encounters node j at time $t_c$ and leaves node j at time $t_s$, we define $t_c - t_s$ as the contact duration of node i for node j.

### 2.1. Social Network Based on Contact History

We assume every node in DTNs records the contact duration information when it encounters every other node, i.e., in a DTN with $n$ nodes, every node should keep contact history for other *n-1* nodes in its buffer. Each node allocates *contact windows* in its buffer, whose size is called *contact window size* $W_s$ which is predefined by the node. Within the range $W_s$ a node keeps its contact duration for other node it has encountered respectively. As time passed the contact window slides by one-unit time, e.g., one second, to keep the recorded contact information latest. The weight of the link between node i and node j at time $t$, $w_{i,j}$, is defined by the following formula:

$$w_{i,j} = \frac{W_s}{\int_{t=0}^{W_s} f(t)dt},$$

where f(t) means the remaining time to the first encounter of the node j after time $t$.

We consider that node i and node j have a stable relationship if the weight $w_{i,j}$ is larger than the predefined threshold $T_h$; an edge between node i and node j is created in two nodes' social networks $SN_i$ and $SN_j$ in this case. Also, node i is added to $SN_j$ and node j is added to $SN_i$. Node i and node j exchange its social network information with each other through hello messages periodically. The algorithm 1 represents the social network construction procedure, where $V_i$ and $E_i$ means set of nodes and edges in $SN_i$, and $N_i$ means set of 1-hop neighbour nodes of node i.

**Algorithm 1** Social Network Construction Procedure For Node i

1: Slide its contact window by a unit time
2: **for** a node j which the node i has at least one contact with **do**
3:     **if** $W_{i,j} > T_h$ **then**
4:         $V_i \cup \{j\}$, $E_i \cup \{(i, j)\}$
5:         **if** $N_j \neq \emptyset$ **then**
6:             $V_i = V_i \cup N_j$
7:             $E_i = E_i \cup \{(j, k)\}$ **for** $k \in N_j$
8:         **end if**
9:     **else**
10:        **if** $j \in V_i$ **then**
11:            $V_i = V_i - \{j\} - (N_j - N_i)$
12:            $E_i = E_i - \{(i, j)\}$, $E_i = E_i - \{(j, k)\}$ **for** $k \in N_j$
13:        **end if**
14:     **end if**
15: **end for**

### 2.2. Expanded Ego Betweenness Centrality

Centrality is a core concept for the analysis of social networks, and betweenness is one of the most prominent measures of centrality. It was introduced independently by Anthonisse (1971) and Freeman (1977), and measures the degree to which a vertex is in a position of brokerage by summing up the fractions of shortest paths between other pairs of vertices that pass through it [6]. We use the betweenness centrality to increase routing efficiency of DTNs. For node i in network, betweenness centrality $C_{B(i)}$ is defined as follows:

$$C_{B(i)} = \sum_{s \neq i \neq t \in V, s < t} \frac{\rho_{st}(i)}{\rho_{st}},$$

where V is the set of nodes in network, $\rho_{st}$ is the number of shortest paths between the node s and t, and $\rho_{st}(i)$ is the number of those shortest paths that include the node i. Due to lack of the whole network-wide end-to-end connectivity, it is hard to utilize the betweenness centrality in DTNs, which requires the entire network information. However, we can use the expanded ego betweenness which is calculated only with the local information [7].

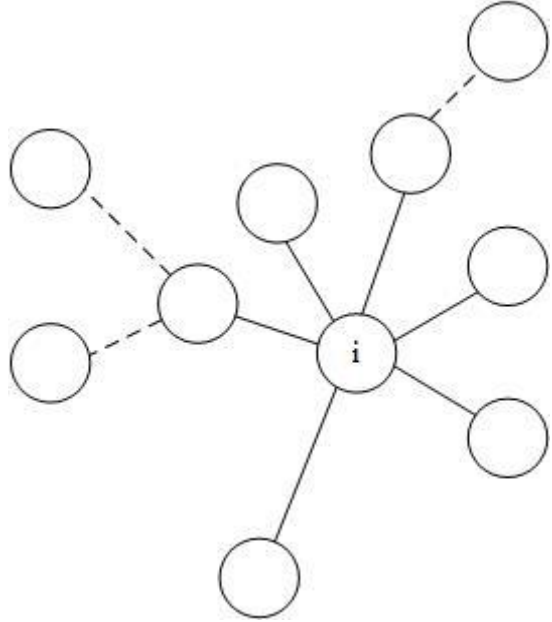

Figure 1. Expanded Ego Network

Figure 1 illustrates an expanded ego network of node i, which is constituted by the ego (node i), its 1-hop neighbours and its 2-hop neighbours. While solid links presents the ego network introduced in [8,9], the entire expanded ego network (Figure 1) is constructed by algorithm 1. Y.H. Kim et al. verified that the expanded ego betweenness centrality is highly correlated with the betweenness centrality in a complete network [7]. Hence we use expanded ego betweenness centrality instead in subsequent studies.

### 2.3. Endpoint Biased Expanded Ego Betweenness Centrality

In some cases, it may be inappropriate to have pairs of vertices depend on intermediaries, but not on themselves (or at least one of them as the source or target). In an information exchange network, for instance, one might argue that the source of information has just as much control over its content as anyone passing it on [6]. So that we have the following formula:

$$C_{EB(i)} = \sum_{s<t} \frac{\rho_{st}(i)}{\rho_{st}}$$

For the above mentioned reasons this routing strategy should also be used in ego betweenness networks. The following section will illustrate its performance.

## 3. ROUTING STRATEGY

In our algorithm, the following steps are required for a node before making routing decisions:

1) Calculate the link weight based on the contact history recorded for other encountered nodes.

2) Construct the virtual social network.

3) Calculate the expanded ego betweenness centrality.

Then a node can make decision on a relay node when it tries send messages to the destination.

### 3.1. Link Weight Based Strategy

Imagine the following scenario: node i tries to send a message to node k and it just contacts with node j; now node i has to make a decision if it should forward the message to node j. At this time, node i compares two link weight values $w_{i,k}$ and $w_{j,k}$. Higher link weight value indicates the corresponding two nodes are more friendly and it is more possible for these two nodes to meet again. Node i forwards the data packet to node j if the following condition is met.

$$\text{Condition I: } w_{i,k} > w_{j,k}$$

### 3.2. Expanded Ego Betweenness Centrality Based Strategy

We also use expanded ego betweenness centrality to increase the overall routing efficiency. In DTNs some nodes may have quite low link weights for they hardly meet other nodes. Using link weight based strategy some nodes may not find a proper relay if the destination node is one of these isolated nodes. Under the circumstances the data packet could be remove from the sender's buffer by TTL expiration before delivered to the destination. To prevent this situation node j sends the message to node j if $C_{B(j)}$ is larger than $C_{B(i)}$ even though Condition I is not met. The second condition for message forwarding is as follow.

$$\text{Condition II: } C_{B(j)} > C_{B(i)}$$

### 3.3. Endpoint Biased Expanded Ego Betweenness Centrality Based Strategy

For the same reasons discussed in previous sub-section, node j sends the message to node j if $C_{EB(j)}$ is larger than $C_{EB(i)}$ even though Condition I is not met, if network adopts the modified version (endpoint biased) of expanded ego betweenness centrality based strategy. So we have the Condition III:

$$\text{Condition III: } C_{EB(j)} > C_{EB(i)}$$

In later sections we will also focus on the performance differences of strategies mentioned in sub-sections 3.2 and 3.3.

### 3.4. Message Delivery Cost Reduction Mechanism

This sub-section proposes a mechanism to reduce the node's buffer consumption. If there is a data packet in node i's buffer, which is destined to send to node k, and node i just contact with node j which satisfies $w_{j,k} > w_{i,k}$, node i will forward the message to node j. Under the circumstances if $w_{j,k}$ is the largest among the link weights between other nodes in $SN_i$, node i deletes this data packet after forwarding the message to prevent further dissemination. Algorithm 2 illustrates in detail this message delivery cost reduction mechanism.

**Algorithm 2** Message Forwarding Procedure

1: **upon** reception of a Hello message from a node $j$ **do**
2: **if** $j \in SN_i$ **then**
3:    **if** $w_{j,k} > w_{i,k}$ **then**
4:       forward the message destined for the node $k$ to the node $j$
5:       **if** $w_{j,k} > w_{m,k}$ **for** all m $\in SN_i$ **then**
6:          delete the message destined for the node $k$ from the node $i$'s buffer
7:       **end if**
8:    **else if** $C_B(j) > C_B(i)$ **then**
9:       forward the message destined for the node $k$ to the node $j$
10:   **end if**
11: **end if**

## 4. PERFORMANCE EVALUATION

In this section, we provide simulation results for two proposed schemes, and compare them with the epidemic [10] and the friendship [5] routing schemes. Proposed Algorithm I refers to the scheme based on link weight and expanded ego betweenness centrality; Proposed Algorithm II is the scheme based on link weight and endpoint biased expanded ego betweenness centrality. The following metrics are used to measure the performance of various schemes:

1) Message delivery ratio.

2) Message delivery cost.

3) Message delivery efficiency.

The message delivery ratio is the proportion of messages that destinations successfully received among the total messages generated by source nodes. The delivery cost is the average number of forwards done during the simulation. The delivery efficiency is defined as the ratio of delivery ratio to the delivery cost.

Our simulation is based on predefined node mobility data; all nodes in the network record contact information during the simulation. We generated 1000 messages, each from a node to a random destination after $W_s$ period of time. All the data packets are given a certain TTL value; any TTL-expired packet will be removed from buffer. The simulation ended when 1000 messages are delivered successfully or expired. All results are averages over 10 runs. Table 1 gives the parameters.

Table 1. Simulation Parameters

| Area Size(m) | 1000 * 1500 |
|---|---|
| Number of Nodes | 25, 75 |
| Communication Range (m) | 3 |
| Moving Speed (m/s) | 0.5, 1.0, 1.25, 1.5 |
| Contact Window Size ($W_s$, second) | 600 |
| TTL (second) | 60, 120, 180, 240, 300, 360 |
| Threshold (Th) | 0.01 |
| Number of Messages | 1000 |

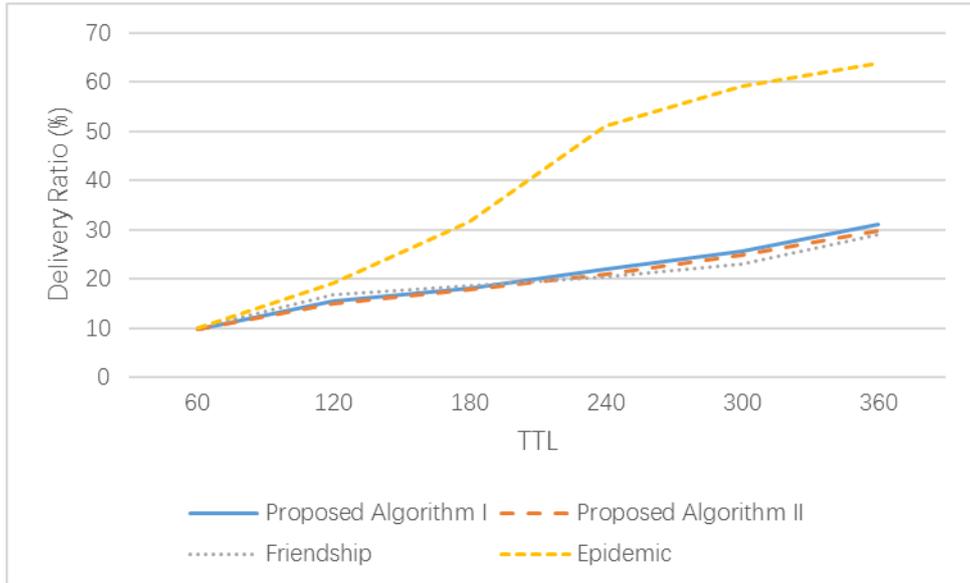

(a) N = 25

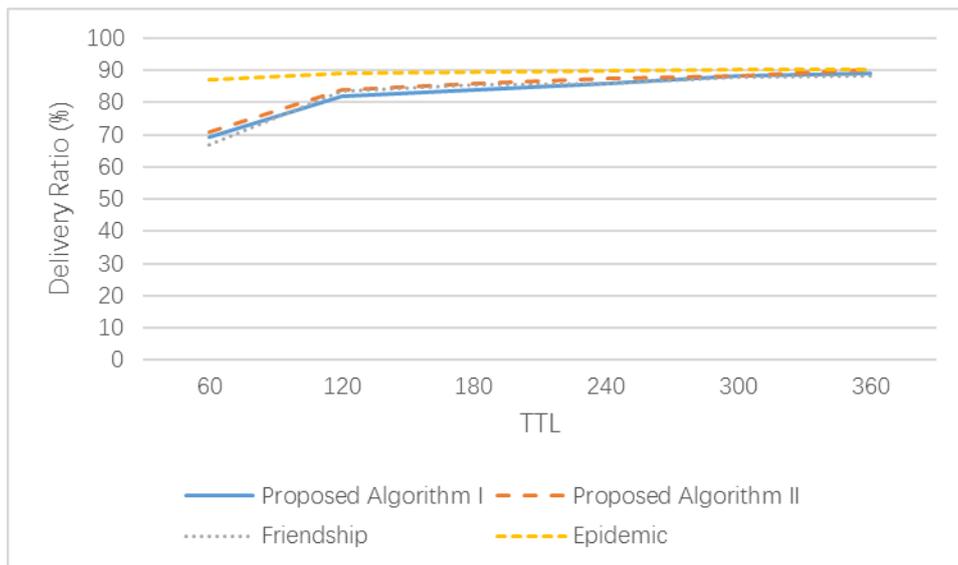

(b) N = 75

Figure 2. Delivery Ratio

Figure 2 shows the delivery ratios achieved by each routing schemes. As TTL increases, all the schemes tend to have higher success rate of delivering. Epidemic routing scheme has the highest delivery ratio as expected. It is noteworthy that the rest schemes have similar performance.

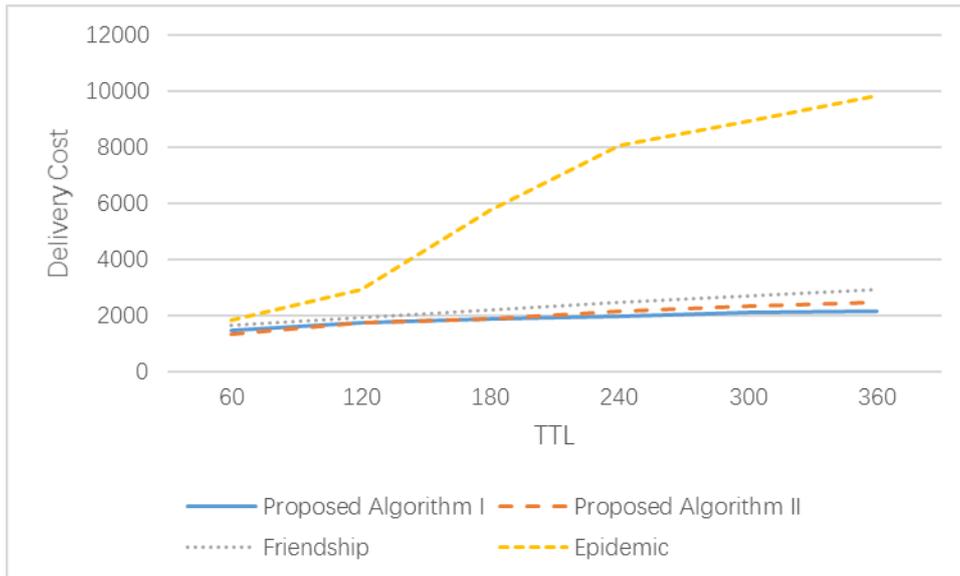

(a) N = 25

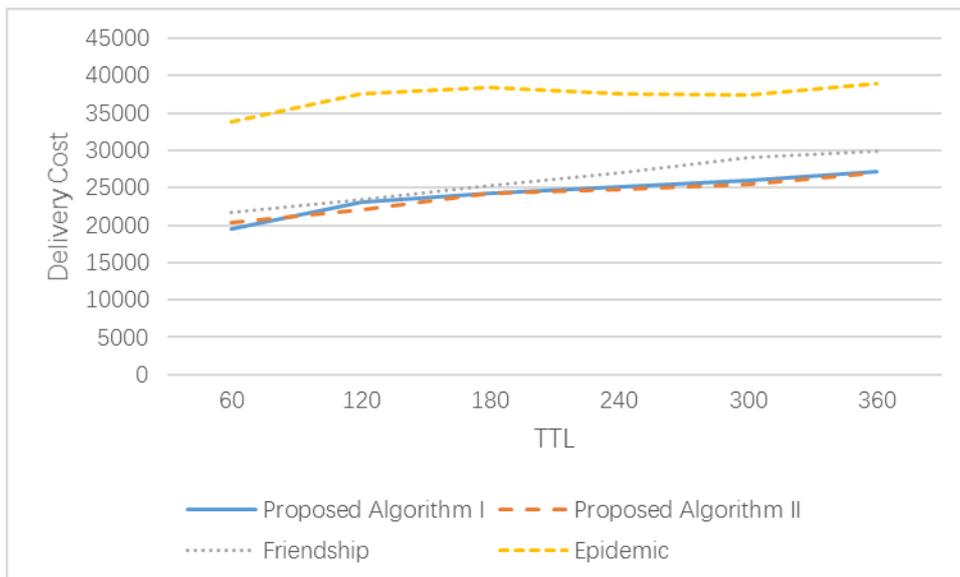

(b) N = 75

Figure 3. Delivery Cost

Figure 3 shows the delivery cost achieved by each schemes. Epidemic routing scheme has the worst performance for the delivery cost because it uses flooding strategy. We can also see both two proposed schemes perform better than friendship routing scheme.

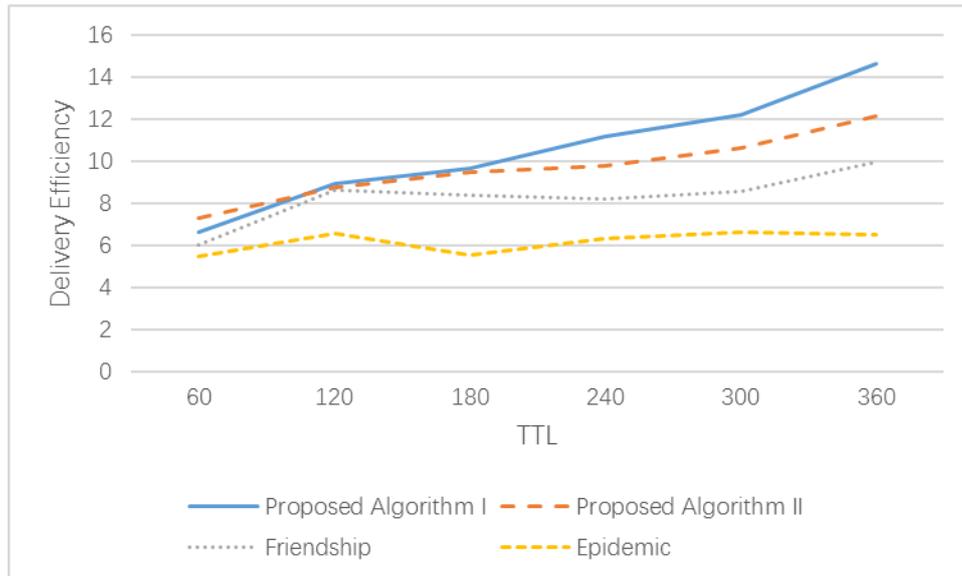

(a) N = 25

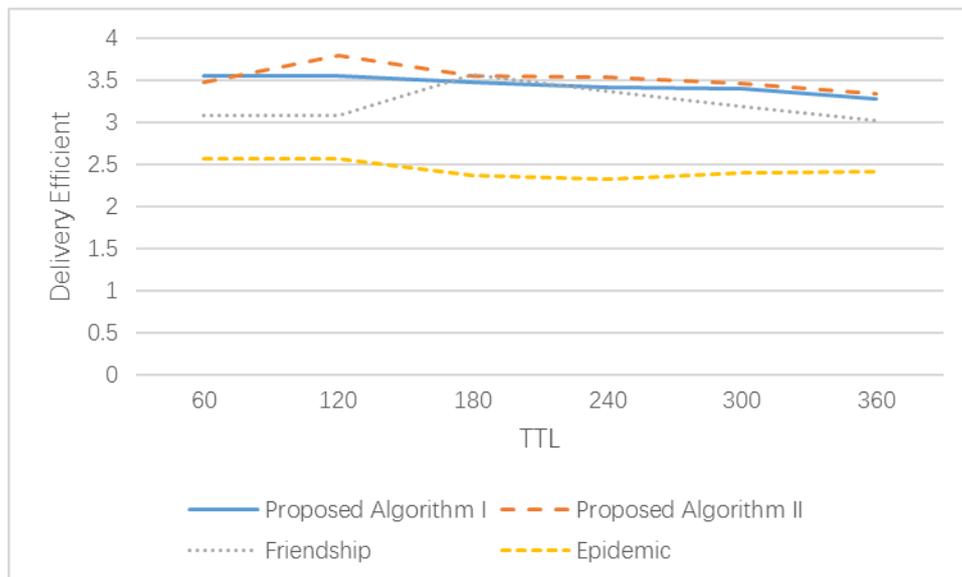

(b) N = 75

Figure 4. Delivery Efficiency

Figure 4 shows the delivery efficiency achieved by each schemes. We can see two proposed schemes have better routing efficiency than epidemic and friendship routing schemes. Compared to the friendship routing schemes, our schemes have higher efficiency by reducing the delivery cost with similar delivery ratio. It is also noted that Proposed Algorithm I performs slightly better than Proposed Algorithm II in a small DTN (N = 25).

## 5. CONCLUSIONS

In this paper we introduced a routing mechanism and its variation in delay-tolerant networks, in which each node forwards their messages to nodes that contain the destination node in their friendship communities. We evaluated our schemes through simulations using predefined node mobility data, and demonstrated that our schemes achieved higher efficiency than two routing schemes proposed previously.

## ACKNOWLEDGEMENTS

The main research is supported by National Natural Science Foundation of China.